\documentclass[aip,jcp,reprint,showpacs,longbibliography,noeprint]{revtex4-1} 
\usepackage[utf8]{inputenc}
\usepackage[english]{babel}
\usepackage{cmap} 
\usepackage[T1]{fontenc} 

\usepackage{amsmath,amssymb,bm}
\IfFileExists{mtpro2.sty}
{ 
  \usepackage{times}
  \usepackage[lite,eucal,subscriptcorrection,slantedGreek,zswash]{mtpro2}
}{
\usepackage{txfonts}

}

\usepackage{mleftright}
\usepackage{microtype}
\usepackage{bbm}
\usepackage[title]{appendix}

\usepackage{graphicx,grffile,subfigure}
\graphicspath{{figures/}}
\newlength{\figwidth}
\usepackage{url,hyperref}
\usepackage[table,usenames,dvipsnames]{xcolor}
\hypersetup{colorlinks=true, linkcolor=BrickRed,
  urlcolor=blue!50!black, citecolor=blue!50!black}

\usepackage[capitalize]{cleveref}
\crefrangelabelformat{equation}{(#3#1#4)$-$(#5#2#6)}
\let\ref\cref

\usepackage{textcomp}

\usepackage{tikz}

\newenvironment{frcseries}{\fontfamily{frc}\selectfont}{}

 \newcommand{\dotdashedline}[1][fill=black]{\tikz 
 [line width=.3ex]
 \draw [#1] [dash pattern=on 3pt off 0.5pt on \the\pgflinewidth off 0.5pt] 
 (0,0) -- (.835,0);}%

\newcommand{\THBM}{T_\mathrm{HBM}}
\newcommand{\THBMk}{T_\mathrm{HBM}^{\mathbf{v}}}
\newcommand{\THBMc}{T_\mathrm{HBM}^{\mathbf{x}}}

\newcommand{\zetaHBM}{\zeta_\mathrm{HBM}}

\renewcommand{\vec}[1]{\mathbf{#1}}
\renewcommand{\tensor}[1]{\underline{\underline{#1}}}

\newcommand{\kb}{k_{\rm B}}
\renewcommand{\vec}[1]{\mathbf{#1}}
\newcommand{\upd}{\mathrm{d}}

\newcommand{\complexi}{\mathbbm{i}}
\newcommand{\imaginary}{\operatorname{\mathbb{I}m}}
\newcommand{\real}{\operatorname{\mathbb{R}e}}

\definecolor{myblue}{RGB}{94.3418,129.735, 181.708}
\definecolor{myokker}{RGB}{225.465, 156.426, 36.3651}
\definecolor{mygreen}{RGB}{143.406, 177.042, 49.8906}
\definecolor{myorange}{RGB}{236.167, 98.7203, 53.5498}
\definecolor{mypurple}{RGB}{135.293, 120.48, 179.546}
\definecolor{mybrown}{RGB}{197.652, 110.478, 26.2111}
\definecolor{mycyan}{RGB}{93.1579, 158.336, 200.281}
\definecolor{myyellow}{RGB}{255, 192, 0}
\definecolor{mymagenta}{RGB}{165.792, 96.809, 157.193}

\usepackage[normalem]{ulem}

\begin{document}

\title{The effective temperature for the thermal fluctuations in hot Brownian motion}

\author{Mayank Srivastava}
\affiliation{
  Indian Institute of Science Education and Research Mohali,
  Knowledge City, Sector 81, S. A. S. Nagar,
  Manauli-140306, India
}

\author{Dipanjan Chakraborty}
\email{chakraborty@iisermohali.ac.in}
\affiliation{
  Indian Institute of Science Education and Research Mohali,
  Knowledge City, Sector 81, S. A. S. Nagar,
  Manauli-140306, India
}

\date{\today}
\begin{abstract} \noindent We revisit the effective parameter
  description of hot Brownian motion -- a scenario where a colloidal
  particle is kept at an elevated temperature than the ambient
  fluid. Due to the time scale separation between heat diffusion and
  particle motion, a stationary halo of hot fluid is carried along
  with the particle, resulting in a spatially varying comoving
  temperature and viscosity profile. The resultant Brownian motion in
  the overdamped limit can be well described by a Langevin equation
  with effective parameters such as effective temperature $\THBM$ and
  friction coefficient $\zetaHBM$ that quantifies the thermal
  fluctuations and the diffusivity of the particle. These parameters
  can exactly be calculated using the framework of fluctuating
  hydrodynamics and requires the knowledge of the complete flow field
  and the temperature field around the particle. Additionally, it was
  also observed that configurational and the kinetic degrees of
  freedom admits to different effective temperatures, $\THBMc$ and
  $\THBMk$, respectively, with the former predicted accurately from
  fluctuating hydrodynamics. A more rigorous calculation by Falasco
  et. al. \emph{Physical Review E} , \textbf{90}, $032131(2014)$
  extends the overdamped description to a generalized Langevin
  equation where the effective temperature becomes frequency dependent
  and consequently, for any temperature measurement from a Brownian
  trajectory requires the knowledge of this frequency dependence. We
  use this framework to expand on this earlier work and look at the
  first order correction to the limiting values in the hydrodynamic
  limit and the kinetic limit. We use the linearized Stokes equation
  and a constant viscosity approximation to calculate the dissipation
  function in the fluid. The effective temperature is calculated from
  the weighted average of the temperature field with the dissipation
  function. Further, we provide a closed form analytical result for
  effective temperature in the small as well high frequency limit. Since
  hot Brownian motion can be used to probe the local environment in
  complex systems, we have also calculated the effective diffusivity
  of the particle in the small frequency limit. To look into the
  kinetic temperature, the velocity autocorrelation function is
  computed from the generalized Langevin equation and the
  Wiener-Khinchine theorem and numerically integrated to evaluate
  $\THBMk$ as a function of the ratio of particle density and fluid
  density $\rho_P/\rho_0$. The two limiting cases of
  $\rho_P/\rho_0 \to 0 $ and $\rho_P/\rho_0 \to \infty $ is also
  discussed.

\end{abstract}

\pacs{...}

\maketitle

\section{Introduction}

Nonequilibrium scenarios of an equilibrium physical phenomena often
serve as a testing field for the equilibrium theories and their limits of
validity. Often, the description of such a nonequilibrium counterpart
is done within the framework of the equilibrium physics and entails
the description using renormalized parameters for the equilibrium
theory.  Perhaps, the most popular system which has garnered an
unfettered attention from the scientific community is that of Brownian
motion, due to its ubiquitous existence in the mesoscopic
world and its nonequilibrium generalization. One such scenario is that of 
hot Brownian motion \cite{Rings:2010iy,Rings:2011gj,Chakraborty:2011kk}.
In hot Brownian motion a colloid is kept at a higher temperature than the 
ambient fluid. Since heat diffusion happens at faster time scale as 
compared to the particle diffusion, the heating of the colloid results in a
stationary temperature profile which is comoving with the colloid
itself. In a fluid, the resulting spatially varying temperature
profile leads to a spatially varying viscosity profile. The ensuing
motion is still diffusive, but with an enhanced diffusion coefficient.
An attempt to describe this enhanced diffusive motion using the coarse
grained Langevin picture requires the use of renormalized parameters
to quantify the thermal fluctuations in the system. Even though these
parameters can be viewed as phenomenological quantities that is used
to simplify the underlying dynamics, nevertheless they can be
rigorously estimated from the hydrodynamical description of the
motion \cite{Rings:2010iy,Rings:2011gj}. 

The cost that we pay in having an effective description is that there
is not one single temperature that quantifies the thermal
fluctuations. The different degrees of motion, as well as different
modes of motion have different effective temperature. Consequently, a
set of these effective temperature $\THBM$ is used for describing the
rotational and translational motion of a heated colloid, as well as
for the configurational and kinetic degrees of freedom for the two
modes of motion. Consequently, $\THBMc$ enters the generalized Stokes
Einstein relation and the Boltzmann distribution in the presence of an
external potential $U(\mathbf{x})$:\cite{Chakraborty:2011kk}
\begin{equation}
  \label{eq:GSER_BZ}
  D_{\rm HBM}=\kb \THBMc/\zeta_{\rm HBM} \quad \text{and} \quad \mathcal{P}(\mathbf{x})
  \sim e^{-U(\mathbf{x})/\kb \THBMc}
\end{equation}
and $\THBMk$ defines the width of the velocity fluctuations
$\mathbf{V} $ of the heated colloid with mass $M$,
\begin{equation}
  \label{eq:THBMk}
  \quad \mathcal{P}(\mathbf{V})
  \sim e^{-M\mathbf{V}^2/2 \kb \THBMk} \quad \text{and} \quad \langle
  \vec{V}^2 \rangle =\frac{3 \kb \THBMk}{M}
\end{equation}
The more general approach is to interpret this effective temperature
as a frequency dependent quantity $\mathcal{T}(\omega)$ that together
with Basset-Boussinesq force quantifies the thermal fluctuations
acting in the colloid. 

In this article we take a look at this effective parameter
$\mathcal{T}(\omega)$ and using the framework of fluctuating
hydrodynamics developed earlier\cite{Chakraborty:2011kk} we derive the
limiting behavior of the quantity $\mathcal{T}(\omega)$ for an
incompressible solvent. In the earlier work by Falsaco et. al
\cite{Falasco:2014iq}, the overdamped description of hot Brownian
motion\cite{Chakraborty:2011kk} was formally extended to it's
generalized form 
\begin{equation}
  \label{eq:GLE}
  \dot{\vec{P}}=-\frac{1}{M}\int_0^t \zeta(t-t') \vec{P}(t')+ \boldsymbol{\xi}
\end{equation}
with the effective parameters inheriting the frequency dependence.The non-equilibrium thermal fluctuations
for the Brownian particle can now be written as :
\begin{equation}
\langle \xi_i(\omega)\xi_j(\omega')\rangle=2k_B
\mathcal{T}(\omega)\operatorname{\mathbb{R}e}[\zeta(\omega)]\delta_{ij}\delta(\omega+\omega')
\label{eq:fdt_Brownian_particle}
\end{equation}
where $\zeta(\omega)$ is the effective friction coefficient
which can be determined from the solution to the Stokes equation.

The current work is motivated by the fact that the limiting behavior
of the frequency dependent temperature can be used in
\emph{nonequilibrium thermometry}\cite{Wulfert2017}. Furthermore, in
experiments as well as in computer simulations the finite time
resolution puts an upper cutoff on the frequency of sampling and
therefore the correct limiting values are not accessible. Hence, it
becomes essential to have a closed form analytic expression for the
effective temperature in the limit of $\omega \to 0$ and $\infty$.
Using the framework of hot Brownian motion, we calculate the frequency
dependent dissipation function in a closed form and derive the
expression of $\mathcal{T}(\omega)$ and the frequency dependent
diffusivity $D(\omega)$. The expressions for these effective
parameters, in particular the first order correction to
$\mathcal{T}(\omega)$ and $D(\omega)$ near $\omega \to 0$ are
important in validating the coarse grained dynamics of the heated
particle and its application as a spectroscopic tool. We note that the
limit of $\omega \to \infty$ has the physical interpretation of
frequencies larger than the inverse of the vorticity diffusion time.

We summarize the main results of our work in the following lines. The
expression for the effective temperature quantifying the thermal
fluctuations is given by the expression
\begin{equation}
  \label{eq:effective_noise_temperature}
  \mathcal{T}(\omega)=T_0+\frac{5}{12} \Delta T
  \frac{\mathcal{F}(\omega)}{(1+ a  \sqrt{\frac{\rho_0 \omega}{2 \eta}} )}
\end{equation}
which in the limit of $\omega \to 0$ yields 
\begin{equation}
  \label{eq:effective_noise_temp_low_freq}
\mathcal{T}(\omega) \approx T_0+\frac{5 \Delta T}{12} \left (1+ a  \sqrt{\frac{\rho_0 \omega}{2 \eta}}
   \right).
\end{equation}. On the opposite limit of $\omega \to \infty$,
$\mathcal{T}(\omega)$ take the form 
\begin{equation}
  \label{eq:eff_temp_high_freq_summary}
\mathcal{T}(\omega) \approx T_0+\Delta T \left(
  1-\frac{1}{3a}\sqrt{\frac{2 \eta}{\rho_0 \omega}} \right)
\end{equation}.

The frequency dependent diffusion coefficient is then evaluated near
$\omega \to 0$ to yield
\begin{equation}
  \label{eq:eff_diffusion_coefficient_smal_frequency}
\begin{split}
   D(\omega) \approx \frac{\kb T_0}{\zeta_0} \bigg[
  &\left(1 +\frac{5}{12} \frac{\Delta T}{T_0} \right)- \\
 & a  \sqrt{\frac{\rho_0 \omega}{2 \eta}}
  \left(1+\frac{5}{6} \frac{\Delta T}{T_0} a  \sqrt{\frac{\rho_0
        \omega}{2 \eta}} \right) \bigg].
\end{split}
\end{equation}

The rest of the article is organized in the following sections. In
\cref{sec:hydrodynamics}, we briefly describe the route to calculate
the effective temperature $\mathcal{T}(\omega)$. In
 \cref{sec:dissipation_function}, we evaluate the complete
frequency dependent dissipation function for an incompressible
solvent. Using the form of this dissipation function, we calculate the
effective temperature for the thermal fluctuations in
\cref{sec:effective_temperature} and the first order correction to its
limiting value as $\omega \to 0$ in
\cref{ssec:low_frequency_limit}. In order to make our results more
transparent in the limit of $\omega \to \infty$ , we approximate the
flow field and the dissipation function in the high frequency regime
and estimate the asymptotic behavior of the effective temperature as
$\omega \to \infty$ in \cref{ssec:high_frequency_limit}. In
\cref{sec:kinetic_temperature} we evaluate the kinetic temperature
$\THBMk$ as a function of the ratio of the density of the particle to
that of the fluid and the effective frequency dependent diffusivity in
\cref{sec:diffusivity}.  Finally, we briefly discuss the significance
of our results in \cref{sec:conclusion}.

\section{Hydrodynamics}
\label{sec:hydrodynamics}
In a hydrodynamic model of Brownian motion, the fluid is usually
treated as a continuum that obeys the linearized Stokes equation
\begin{equation}
  \label{eq:stokes_eqn}
\rho \partial_t \mathbf{u}=\nabla \cdot \tensor{\Pi}
\end{equation}
where the isotropic part of the stress tensor $\tensor{\Pi}$ is given
by the pressure and the deviatoric part is given by the strain
\begin{equation}
  \label{eq:stress}
  \Pi_{ij}= -p \delta_{ij}+ \tau_{ij} \quad \text{with} \quad \tau_{ij}=2\eta \epsilon_{ij},
\end{equation}
where $\epsilon_{ij}=(\partial_j u_i+\partial_i u_j)/2$. The notation
$\partial_i$ and $\partial_t$ denotes a partial derivative with
respect to the spatial coordinate $x_i$ and time $t$, respectively. The
dynamics of the colloid is coupled to the fluid via the boundary
condition on its surface. In what follows, we assume this to be a
perfect no-slip boundary condition. Far away from the colloid the
fluid is assumed stationary. The linearized Stokes equation is usually
further augmented by the incompressibility condition
$\nabla \cdot \mathbf{u}=0$, and is typically valid in the limit of
$\omega \to 0$. \cref{eq:stokes_eqn} can be exactly solved for the
systematic part of the flow field for any arbitrary velocity $U(t)$
prescribed on the surface $\mathcal{S}$ of the colloid. In order to
include thermal fluctuations in the system, the hydrodynamical fields
are split into a systematic and a fluctuating part,
$\vec{u} \to \vec{u}+\vec{\tilde{u}}$, $p \to p+\tilde{p}$ and
$\tensor{\Pi} \to \tensor{\Pi}+\tensor{\tilde{\Pi}}$. The linearity of
\cref{eq:stokes_eqn} ensures that the systematic and the fluctuating
parts decouple. The fluctuating flow field
$\vec{\tilde{u}}(\vec{r},t)$ is driven by the fluctuating stresses
$\tilde{\tau}_{ij}$.  The strength of the fluctuations is determined
by the viscosity of the solvent and the temperature via the
fluctuation--dissipation relation (FDT). In an incompressible solvent
the FDT reads as
\begin{equation}
  \langle \tilde{\tau}_{ij}(\vec{r},t)
  \tilde{\tau}_{kl}(\vec{r'},t')\rangle=2 \eta \kb T
  \delta(\vec{r}-\vec{r'})
  \delta(t-t')\left[\delta_{ij}\delta_{kl}+\delta_{il}\delta_{jk}
  \right]
\label{eq:fdt_stress}
\end{equation}
In hot Brownian motion
\cite{Chakraborty:2011kk,Rings:2011gj,Falasco:2014iq}, even though
there exists a radial temperature and viscosity field due to the
heating of the colloid, the local thermal equilibrium in the fluid
allows the modification of the FDT to include the local temperature
$T(\vec{r})$ and the viscosity $\eta(\vec{r})$ to quantify the
strength of the fluctuating stresses. The solvent degrees of freedom
can now be contracted to yield the equation of motion for the
colloid. A rigorous calculation then yields the expression for the
effective parameters of the coarse grained theory. In particular, the
effective temperature that quantifies the thermal fluctuations for the
coarse grained Langevin dynamics is given by weighted average of
$T(\vec{r},t)$ \cite{Falasco:2014iq} 
\begin{equation}
 \mathcal{T}(\omega)=\frac{\int_V T(\vec{r},\omega)
  \phi(\vec{r},\omega) \upd \vec{r}}{\int_V \phi(\vec{r},\omega) \upd
  \vec{r}}
\label{eq:eff_temp}
\end{equation}
where  $\phi(\vec{r},t)$ is the dissipation function in the fluid,
which for an incompressible solvent reads:
\begin{equation}
\label{eq:dissipation_function}
\phi(\vec{r})=\frac{\eta}{2} \left[ \nabla \vec{u}(\vec{r})+ (\nabla
  \vec{u}(\vec{r}))^\intercal\right]
:\left[ \nabla \vec{u}^{*}(\vec{r})+ (\nabla \vec{u}^{*}(\vec{r}))^\intercal\right],
\end{equation}
where $\vec{u}^{*}$ denotes the complex conjugate of $\vec{u}$ and $(\nabla
  \vec{u}(\vec{r}))^\intercal$ is the transpose of $\nabla
  \vec{u}(\vec{r})$.
The flow field that enters the expression for the dissipation
function, is the systematic part of the flow and obeys the linearized
Stokes \cref{eq:stokes_eqn}.

\section{Dissipation Function}
\label{sec:dissipation_function}

We first turn our attention to calculate the frequency dependent
dissipation function $\phi(\vec{r},\omega)$.
We consider a spherical colloid of radius $a$ suspended in a solvent
of density $\rho_0$, viscosity $\eta$ and at a temperature $T_0$. The
colloid is now heated which results in a radially varying temperature
and viscosity field in the solvent. In a comoving frame of the
particle, the steady state temperature profile $T(r)$ follows the
stationary heat equation
\begin{equation}
\nabla^2 T(\vec{r})= 0
\label{eq:heat_eqn}
\end{equation}
together with the boundary conditions
\begin{equation} 
T(r=a)= T_0+\Delta T \quad \textrm{and} \quad 
T(r\rightarrow \infty)= T_0.
\label{eq:bc_heat_eqn}
\end{equation}
The solution of above equation is the radial field:
\begin{equation}
T(r)=T_0+\Delta T\frac{a}{r}
\label{eq:temp_profile}
\end{equation} 

Since the heating of the
colloids results in a radially symmetric temperature field, the
viscosity in the fluid no longer remains constant throughout. This
spatially varying viscosity can be obtained from the equation of state
$\eta(\rho,T)$. To simplify the calculations, we consider the
situation when the heating of the colloid is small so that the spatial
variation of the viscosity can be ignored in the Stokes equation.

Using the continuity equation
$\partial_t \rho +\nabla \cdot \rho \vec{u}=0$, it is easy to deduce that
in the limit of $\omega \to 0$, the solvent can be treated as
incompressible. The perturbations due to sound propagation has a
characteristic frequency $c/a$ and only comes into play in the high
frequency limit.  Writing down the Stokes equation explicitly in the
frequency space
\begin{equation}
\begin{split}
  -\complexi \omega \rho_0 \vec{u} &=-\nabla p +\eta \nabla^2 \vec{u} \\
 \nabla \cdot \vec{u}&=0,
\end{split}
\label{eq:stokes_equation_freq}
\end{equation}
we seek the frequency dependent solution to the above equation subject
to the boundary conditions:
\begin{equation}
\begin{split}
&\vec{u}(\vec{r},t)= \vec{U}(\vec{r},t)+\boldsymbol{\Omega}(\vec{r},t)\times \vec{r}\quad on\quad S  \\
&\lim_{\vec{r}\to\infty}\vec{u}(\vec{r},t) = 0. 
\end{split}
\label{eq:bc_boundary_condition}
\end{equation}
Without loss of generality, exploiting the spherical symmetry of the
problem, we choose $\vec{U}(t)$ along the direction of the $z$-axis,
so that $\vec{U}(t)=U(t) \hat{\vec{z}}$.  The solution to
\cref{eq:stokes_equation_freq} with the boundary conditions
\cref{eq:bc_boundary_condition} can be derived by writing down
$\vec{u}(\vec{r},\omega)$ as
  \begin{equation}
    \label{eq:flow_field_derivation_1}
\vec{u}(\vec{r},\omega)=\nabla_r \times \nabla_r (h(r) U_\omega).
  \end{equation}
The scalar function $h(r)$ satisfies the differential equation
\begin{equation}
   \label{eq:flow_field_derivation_2}
\nabla_r^4 h(r)+\alpha_\omega^2 \nabla_r^2 h(r)=0
\end{equation}
Solving for $h(r)$ and using the boundary conditions from
\eqref{eq:bc_boundary_condition} the flow field takes the form
\begin{equation}
   \label{eq:flow_field_derivation_3}
   \vec{u}(\vec{r},\omega)=-\frac{2 U_\omega}{r} \frac{\partial
     h}{\partial r} \cos \theta \hat{\vec{r}} + \frac{U_\omega}{r}\left( \frac{\partial
       h}{\partial r} + r\frac{\partial^2
       h}{\partial r^2}\right) \sin \theta \hat{\bm{\theta}}
\end{equation}
with $\frac{\partial h}{\partial r}$ given by 
\begin{equation}
   \label{eq:flow_field_derivation_4}
\frac{\partial h}{\partial r}=\frac{3a}{2\alpha^2_\omega r^2}
\bigg[\left(\complexi \alpha_\omega r -1\right) e^{i \alpha_\omega
  (r-a)}- \left(-1+\complexi \alpha_\omega a +\frac{1}{3}
  \alpha^2_\omega a^2 \right) \bigg]
\end{equation}
The final form of the flow field reads:\cite{Chow1973a}
\begin{equation}
  \label{eq:stokes_eqn_sol}
\begin{split}
u_r (\omega)&=\left[ 2 A_\omega\left(-\frac{1}{r^3}+\frac{\complexi
        \alpha_\omega}{r^2}\right) e^{\complexi \alpha_\omega r} - 
    B_\omega \frac{2}{r^3} \right]U_\omega \cos \theta \\
u_\theta(\omega)&=\left[  A_\omega\left(-\frac{1}{r^3}+\frac{\complexi
        \alpha_\omega}{r^2} +\frac{\alpha_\omega^2}{r} \right) e^{\complexi \alpha_\omega r} - 
    B_\omega \frac{1}{r^3} \right]U_\omega \sin \theta, 
\end{split}
\end{equation}
where the frequency dependence of the quantities have been denoted by
the subscript. $\alpha_\omega$ is given by 
\begin{equation}
  \label{eq:alpha_omega}
\alpha^2_\omega = \complexi \frac{ \omega \rho_0}{\eta}.
\end{equation}
In order to satisfy the boundary condition at infinity,
$\alpha_\omega$ is chosen so that $\imaginary(\alpha_\omega) >0$,
\begin{equation}
  \label{eq:real_im_alpha}
\real(\alpha_\omega) = \imaginary (\alpha_\omega)= \sqrt{\frac{\omega
    \rho_0}{2 \eta}}
\end{equation}
$A_\omega$ and $B_\omega$ are determined from the no-slip boundary
condition on the surface: $u_r(r=a)=U_\omega \cos \theta$,
$u_\theta(r=a)=-U_\omega \sin \theta$ and reads
\begin{equation}
  \label{eq:coefficients_A_B}
\begin{split}
A_\omega&=-\frac{3}{2}\left(\frac{a}{\alpha_\omega^2}\right) e^{-\complexi
  \alpha_\omega a} \\
B_\omega &= \frac{1}{2} \left(\frac{a}{\alpha_\omega^2 }\right) \left(3 - 3
  \complexi \alpha_\omega a - \alpha_\omega^2 a^2 \right)
\end{split}
\end{equation}

Using \cref{eq:dissipation_function}, we write down the dissipation
function explicitly in spherical polar coordinates,
\begin{equation}
\label{eq:dissipation_function_spherical}
\begin{split}
\phi(r,\omega)=&2\eta\left[\left(\partial_r u_r \right) \left(\partial _r u_r\right)^*
+\left(\frac{1}{r} \partial_\theta u_\theta +\frac{u_r}{r}\right) 
\left(\frac{1}{r} \partial_\theta u_\theta +\frac{u_r}{r}\right)^* \right.\\
& \left. + \left(\frac{u_r}{r}+\frac{u_\theta \cot \theta}{r}\right)
\left(\frac{u_r}{r}+\frac{u_\theta \cot \theta}{r}\right)^* \right]\\
& +\eta \left[ r\partial_r \left(\frac{u_\theta}{r}\right)+\frac{\partial_\theta
  u_r}{r}\right]\left[r\partial_r \left(\frac{u_\theta}{r}\right)
+\frac{\partial_\theta u_r}{r}\right]^*.
\end{split}
\end{equation}
Plugging in the flow field from \cref{eq:stokes_eqn_sol}, the
dissipation function can be written down in a closed analytical
form. To simplify this expression, we define a frequency dependent
length scale $\lambda_\omega$ as,
\begin{equation}
\label{eq:k_omega}
 \lambda_\omega=k^{-1}_\omega=\sqrt{\frac{2 \,\eta}{\rho_0 \omega}}. 
\end{equation}
Using this definition for $k_\omega$, the dissipation function is
given by:
\begin{equation}
\label{eq:dissipation_function_1}
\begin{split}
  \phi(r,\omega)= 2 \eta |U_\omega|^2\left[ \frac{3}{4} \left(\frac{9 a}{k^2_\omega}
    \right)^2 \right.
    &\frac{1}{r^8} \mathcal{A} \mathcal{A}^* \cos^2 \theta \,
+ \\
  & \left. \frac{1}{4}\left(\frac{9 a}{ k^2_\omega} \right)^2 
    \frac{1}{r^8}\mathcal{B} \mathcal{B}^* \sin^2 \theta \right],
\end{split}
\end{equation} where $\mathcal{A}^*$ and
$\mathcal{B}^*$ are the complex conjugate of $\mathcal{A}$ and
$\mathcal{B}$, respectively. The terms $\mathcal{A}$ and $\mathcal{B}$
are functions of $r$ and $\omega$ and a detailed expression for them
are given in \cref{sec:dissipation_function_detail}.

\section{Effective Noise Temperature}
\label{sec:effective_temperature}
Finally, we are now in a position to calculate the effective frequency
dependent temperature $\mathcal{T}(\omega)$ using
\cref{eq:eff_temp}. The denominator, which is the integral of the
dissipation function over the whole volume yields,
\begin{equation}
  \label{eq:dissipation_integral}
\begin{split}
\int \upd \vec{r} \phi(\vec{r},&\omega)= 2 \eta |U_\omega|^2
\left(\frac{9 a}{k^2_\omega}\right)^2 \int_a^{\infty} r^2 \upd r \int_0^{\pi} \upd \sin
\theta \theta
\int_0^{2 \pi}\upd \phi \\
 & \qquad \qquad \quad \quad \left[\frac{3}{4} \frac{\mathcal{A}\mathcal{A}^*}{r^8} \cos^2 \theta
    + \frac{1}{4} \frac{\mathcal{B}\mathcal{B}^*}{r^8} \sin^2
    \theta\right]\\
&=4\pi \eta |U_\omega|^2
\left(\frac{9 a}{k^2_\omega}\right)^2 \int_a^{\infty} r^2 \upd r   \left[\frac{1}{2} \frac{\mathcal{A}\mathcal{A}^*}{r^8} 
    + \frac{1}{3} \frac{\mathcal{B}\mathcal{B}^*}{r^8}\right]\\
&=4\pi \eta |U_\omega|^2\left(\frac{9 a}{k^2_\omega}\right)^2
\left(\frac{k_\omega^4}{27 a}\right) \left(1+a k_\omega \right)
\end{split}
\end{equation}
Using the final form of the above equation, a little algebra gives the result
\begin{equation}
  \label{eq:dissipation_integral_final}
\int \upd \vec{r} \phi(\vec{r},\omega)=4\pi \eta |U_\omega|^2 3 a \left(1+a k_\omega \right).
\end{equation}
Note that in the limit of $\omega \to 0$ the integral evaluates to
$$\lim_{\omega \to 0} \int \upd \vec{r} \phi(\vec{r},\omega)=4\pi \eta |U_\omega|^2 3 a$$.

Using the form of the temperature profile from \cref{eq:temp_profile},
the numerator in \cref{eq:eff_temp} takes the form 
\begin{equation}
  \int \upd \vec{r} \left(T_0 + \Delta T \frac{a}{r} \right)\phi(\vec{r},\omega)
\end{equation}
The first term in the above equation is trivial and is given by the
spatial integral of the dissipation function rescaled by a factor $T_0$
\begin{equation}
\label{eq:numerator_first_term}
  \int \upd \vec{r} T_0 \phi(\vec{r},\omega)= 4\pi \eta T_0 |U_\omega|^2 3 a \left(1+a k_\omega \right).
\end{equation}
The second term which integrates the spatially varying part of the temperature
field weighted by the dissipation function yields a slightly
complicated expression, 
\begin{equation}
\label{eq:numerator_second_term}
  \Delta T \int \upd \vec{r} \frac{a}{r} \phi(\vec{r},\omega)=4 \pi
  \eta |U_\omega|^2 \Delta T \frac{5a}{4} \mathcal{F}(\omega)
\end{equation}
with $\mathcal{F}(\omega)$ given by 
\begin{equation}
\begin{split}
\label{eq:f_omega}
  \mathcal{F}(\omega)&=\bigg[ \left(1+ 2 a k_\omega +\frac{6}{5} a^2
    k_\omega^2 +\frac{2}{5} a^3 k_\omega^3 \right)\\
& +a^2 k_\omega^2 e^{a
  k_\omega} \bigg( -6 \pi (1+ a k_\omega) \cos a k_\omega   \\
&  -6 a k_\omega \left( 1+ \frac{2
      \pi}{3} a k_\omega \right) \sin a k_\omega + \bigg(3 \complexi
  (1+ ak_\omega)\\
&+ 3 a k_\omega \left(1 + \frac{2}{3} a k_\omega\right) \bigg)
\mathcal{U}(\omega) \bigg]
\end{split}
\end{equation}
with
$\mathcal{U}(\omega)= e^{\complexi a k_\omega} Ei(-a\alpha_\omega)
-e^{-\complexi a k_\omega} Ei(-a\alpha^*_\omega)$ and
$Ei(z)=-\int_z^\infty e^{-t} \upd t$.
Using
\cref{eq:eff_temp,eq:dissipation_integral_final,eq:numerator_first_term,eq:numerator_second_term},
the effective frequency dependent temperature $\mathcal{T}(\omega)$
takes the form
\begin{equation}
\label{eq:T_omega}
  \mathcal{T}(\omega)=T_0+\frac{5}{12} \Delta T
  \frac{\mathcal{F}(\omega)}{(1+ a k_\omega)}
\end{equation}

\begin{figure}[!t]
  \centering
  \includegraphics[width=0.8\linewidth]{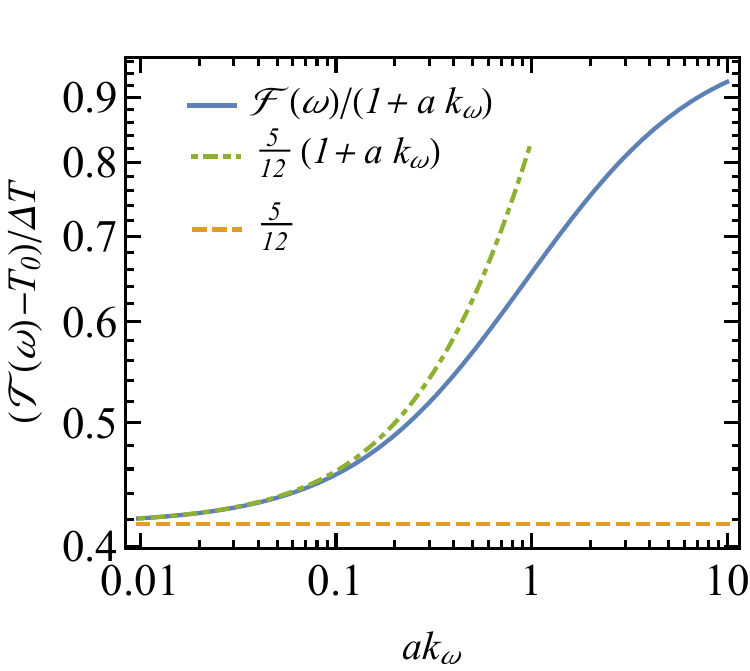}
  \caption{Plot of $(\mathcal{T}(\omega)-T_0)/\Delta T$ using
    \cref{eq:T_omega} as a function of the dimensionless quantity
    $a k_\omega$ (the solid line). The low
    frequency approximation in \eqref{eq:effective_temp_low_freq} is
    depicted by the dot-dashed line 
    while the horizontal dashed line  denotes the
    value of the ratio  $(\mathcal{T}(\omega)-T_0)/\Delta T$ at zero frequency.}
\end{figure}

\subsection{The Low Frequency Limit of $\mathcal{T}(\omega)$}
\label{ssec:low_frequency_limit}
In this section, we estimate the effective temperature
$\mathcal{T}(\omega)$ in the limit of $\omega \to 0$ and derive the
first order correction in $\omega$ to its limiting value of
$\THBMc$. To this end we expand the function $\mathcal{F}(\omega)$ in
Taylor series as $\omega \to 0$. For small values of $\omega$ one can
very easily read off from \cref{eq:f_omega} 
\begin{equation}
  \label{eq:f_omega_small}
\mathcal{F}(\omega) \approx 1+2a k_\omega \quad \text{for $\omega \to 0$.} 
\end{equation}
Hence, the frequency dependence of the effective temperature take the
form
\begin{equation}
  \label{eq:effective_temp_low_freq}
\mathcal{T}(\omega) \approx T_0+\frac{5 \Delta T}{12} \left (1+ a
  k_\omega \right)=T_0+\frac{5 \Delta T}{12} \left (1+ a
  \sqrt{\frac{\rho_0 \omega}{2 \eta}} \right)
\end{equation}
Note  that $\tau_\nu=a^2/\nu=a^2/(\eta/\rho_0)$ is the typical
vorticity diffusion timescale in the fluid.

\subsection{The High Frequency Limit of $\mathcal{T}(\omega)$}
\label{ssec:high_frequency_limit}
The general expression for the effective temperature as given in
\cref{eq:T_omega} is not very transparent and the high frequency
behavior is difficult to extract from this expression, in particular
spurious oscillations appear due to the behavior of the integrals
$Ei(x)$ . It should be pointed out here, that the limit of
$\omega \to \infty$ implies that we look at large frequencies with
$\omega >> \nu/2a^2$ but $\omega < c/a$, where $c$ is the adiabatic
sound speed in the solvent. This allows us to treat the solvent
incompressible and use the Stokes solution as given in
\cref{eq:stokes_eqn_sol}.

In order to have a gainful insight into the effective temperature at
large frequencies, instead of working with \cref{eq:T_omega}, we look
at the flow field given in \cref{eq:stokes_eqn_sol} in the high
frequency limit. Since $\alpha_\omega \sim \sqrt{\omega}$, the terms
with coefficient $A_\omega$ in $u_r$ and $u_\theta$ are exponentially
damped in the expression of dissipation function. A careful
consideration gives the approximate flow field in the high frequency
limit as
\begin{equation}
  \label{eq:flow_field_high_frequency}
\begin{split}
u_r (\omega)&=\left[ 2 A_\omega\left(\frac{\complexi
        \alpha_\omega}{r^2}\right) e^{\complexi \alpha_\omega r} - 
    B_\omega \frac{2}{r^3} \right]U_\omega \cos \theta \\
u_\theta(\omega)&=\left[  A_\omega\left(\frac{\alpha_\omega^2}{r} \right) e^{\complexi \alpha_\omega r} - 
    B_\omega \frac{1}{r^3} \right]U_\omega \sin \theta, 
\end{split}
\end{equation}

\begin{figure}[!t]
\label{fig:flow_field_plots}
  \centering
\includegraphics[width=\linewidth]{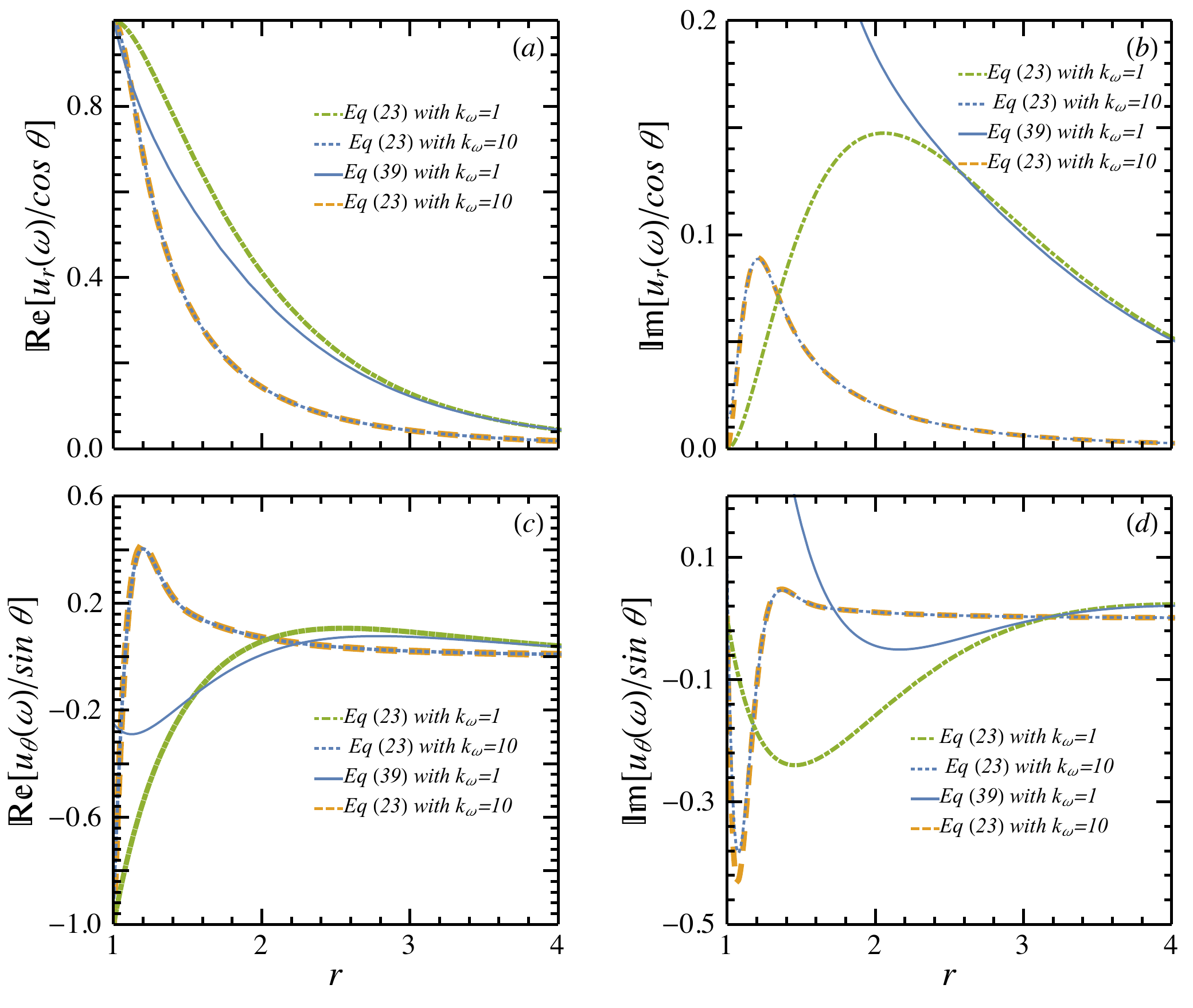}  
\caption{The plots depict the actual flow fields in the solvent as a
  function of $k_\omega$ from \cref{eq:stokes_eqn_sol} and its
  comparison to the high frequency approximation in
  \cref{eq:flow_field_high_frequency}.  Plot of the real and imaginary
  parts of the flow field $u_r(\omega)/cos \theta$ (real part: figure
  (a), imaginary part: figure (b)) and $u_\theta(\omega)/sin \theta$
  (real part: figure (c), imaginary part: figure (d)) using the
  complete solution of the Stokes equation \cref{eq:stokes_eqn_sol}
  for two different values of $k_\omega$: $k_\omega=1$ (the dot-dashed
  line) and $10$ (the dotted line). The corresponding high frequency
  approximations given in \cref{eq:flow_field_high_frequency} is also
  plotted for the same values of $k_\omega$: $k_\omega=1$ (the solid
  line ) and $10$ (the dashed line). In the high frequency limit the
  approximation of given in \cref{eq:flow_field_high_frequency}
  matches with the flow field for all values of $r$ and consequently
  the dotted line  and the dashed line
   overlap to give an appearance
  of a dot-dashed line.The radius of the nanoparticle $a$ is unity. }
\end{figure}

The dissipation function calculated from the above expressions for the
flow field yields:
\begin{equation}
  \label{eq:dissipation_function_high_frequency}
\begin{split}
\phi(\vec{r},\omega)=2 \eta |U_\omega|^2 \left( \frac{81 a^2 \eta^2}{\rho^2_0
    \omega^2}\right)\frac{1}{r^8} \bigg[\mathcal{A}_1 \mathcal{A}^*_1
3\cos^2 \theta +\\
\mathcal{B}_1 \mathcal{B}^*_1
\sin^2 \theta  \bigg]
\end{split}
\end{equation}
with
\begin{equation}
  \label{eq:A_1}
\begin{split}
\mathcal{A}_1=1+\frac{2 \complexi}{3} a^2 k_\omega^2
\left(1-\frac{r^2}{a^2} e^{-(r-a)(1+\complexi)k_\omega}\right)\\
+(1+\complexi) a k_\omega \left(1-\frac{2}{3}\frac{r}{a}
  e^{-(r-a)(1+\complexi)k_\omega} \right)
\end{split}
\end{equation}
and 
\begin{equation}
  \label{eq:B_1}
\begin{split}
\mathcal{B}_1=1+\frac{2 \complexi}{3} a^2 k_\omega^2
\left(1-\frac{r^2}{a^2} e^{-(r-a)(1+\complexi)k_\omega}\right)\\
+(1+\complexi) a k_\omega \left(1-\frac{1}{3}\frac{r}{a}
  e^{-(r-a)(1+\complexi)k_\omega} \right)\\
 -\complexi (1-\complexi) r^3
k^3_\omega e^{-(r-a)k_\omega}
\end{split}
\end{equation}
Since $k_\omega$ diverges as $\sqrt{\omega}$, the exponentials are
damped in the dissipation function and rapidly approach zero as
$\omega \to \infty$. Hence, we approximate $\mathcal{A}_1$ as
\begin{equation}
  \label{eq:A_1_approx}
\begin{split}
\mathcal{A}_1=1+\frac{2 \complexi}{3} a^2 k_\omega^2
+(1+\complexi) a k_\omega 
\end{split}
\end{equation}
and $\mathcal{B}_1$ as
\begin{equation}
  \label{eq:B_1_approx}
\begin{split}
\mathcal{B}_1=1+\frac{2 \complexi}{3} a^2 k_\omega^2
+(1+\complexi) a k_\omega  -\complexi (1-\complexi) r^3
k^3_\omega e^{-(r-a)k_\omega}.
\end{split}
\end{equation}

\begin{figure}[!t]
  \centering
  \includegraphics[width=\linewidth]{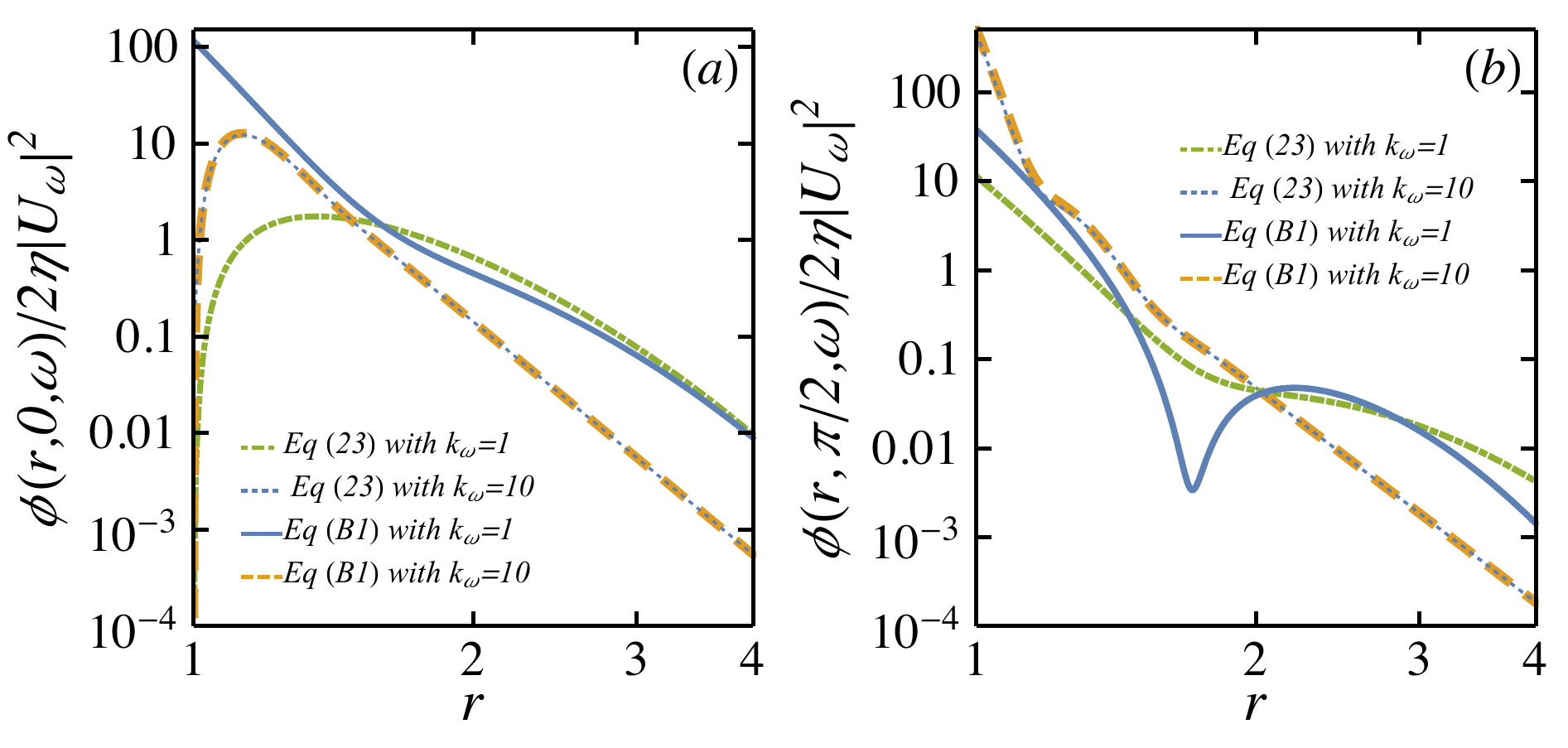}
  \caption{The plots compare the two terms in
    \cref{eq:dissipation_function_1} for the dissipation function
    $\phi(r,\theta,\omega)$ as a function of $k_\omega$ with the high
    frequency approximation in
    \cref{eq:dissipation_function_high_frequency_approx}. Setting the
    value of $\theta=0$ in $\phi(r,\theta,\omega$ picks up the
    pre-factor of the $\cos^2 \theta$ term and $\theta=\pi/2$ picks up
    the pre-factor of the $\sin^2 \theta$.  For smaller values of
    frequency ( $k_\omega=1$), the dissipation function calculated
    from the flow field using \cref{eq:stokes_eqn_sol} (the dot-dashed
    line  matches with that of the high
    frequency approximation (the solid line )
    only asymptotically. On the other hand, for large values of
    frequency ($k_\omega=10$), the dissipation function from the
    Stokes solution (the dotted line ) agrees
    with the high-frequency approximation in
    \cref{eq:dissipation_function_high_frequency_approx} (the dashed
    line) throughout the
    system. 
    As a consequence, the dotted line  and
    the dashed line overlap to give an
    appearance of a dot-dashed line.The radius of the nanoparticle $a$ 
    is taken to be unity.}
\end{figure}
The detailed calculation of the dissipation function, its spatial
integral and the weighted average of $T(\vec{r})$ in the high
frequency limit is provided in \cref{sec:Teff_high_frequency_limit}.
Using \cref{eq:dissipation_function_integral_final} and
\cref{eq:numerator_high_frequency}, the effective temperature in the
high frequency limit reads as
\begin{equation}
\label{eq:eff_temp_high_freq}
  \mathcal{T}(\omega) \approx T_0+\frac{5 \Delta T}{6}\frac{\mathcal{H}(\omega)}{\mathcal{G}(\omega)}
\end{equation}
To get the asymptotic correction to $\mathcal{T}(\omega)$, we look at the series expansion
of the ratio $\mathcal{H}(\omega)/\mathcal{G}(\omega)$ in the limit of
$\omega \to \infty$:
\begin{equation}
\label{eq:series_expansion_H_by_G}
  \frac{\mathcal{H}(x)}{\mathcal{G}(x)} \approx
  \frac{6}{5}-\frac{29}{45}\left(\frac{1}{x} \right)
\end{equation}
Combining the above equation with \cref{eq:eff_temp_high_freq}, we
arrive at the following expression for $\mathcal{T}(\omega)$: 
\begin{equation}
  \label{eq:eff_temp_high_freq_final}
\mathcal{T}(\omega) \approx T_0+\Delta T \left( 1-\frac{29}{90}\frac{1}{a
    k_\omega} \right) \approx T_0+\Delta T \left(
  1-\frac{1}{3a}\sqrt{\frac{2 \eta}{\rho_0 \omega}} \right)
\end{equation}

It should be pointed out that retaining the last terms in both the
equations of \cref{eq:flow_field_high_frequency} leads to an erroneous
result for the kinetic temperature
$\mathcal{T}(\omega \to \infty)=T_0+\frac{5}{6} \Delta T$.

\section{The Kinetic Temperature $\THBMk$}
\label{sec:kinetic_temperature}
Once the complete spectrum of the effective temperature is known to
us, we can use the generalized Langevin equation in \cref{eq:GLE} to
evaluate the velocity autocorrelation function
$\vec{V}(t)\vec{V}(0)$ for the colloid.
The convolution in \cref{eq:GLE} is easier to handle in the frequency
domain and the expression for the velocity of the particle reads as
\begin{equation}
  \label{eq:velocity_colloid}
\vec{V}(\omega)=G(\omega)\bm{\xi}(\omega)
\end{equation}
where the Green's function $G(\omega)$ is given by
\begin{equation}
  \label{eq:greens_function}
G^{-1}(\omega)=\zeta(\omega)-\complexi \omega M
\end{equation}
In order to evaluate the equipartition value $\langle
\vec{V}^2 \rangle$ we use the Wiener-Khinchine theorem 
and write down the velocity autocorrelation function as: 
\begin{equation}
\label{eq:vacf}
  \langle \vec{V}(t) \vec{V}(0) \rangle =\frac{1}{2\pi}\int_{-\infty}^\infty
  \langle \vec{V}(\omega)\cdot \vec{V}(-\omega) \rangle e^{-\complexi
    \omega t} \upd \omega
\end{equation}
From this the equal time correlation reads as 
\begin{equation}
  \label{eq:equal_time_correlation}
  \langle \vec{V}^2 \rangle =\frac{1}{\pi}\int_0^\infty
  |G(\omega)|^2 6\kb \mathcal{T}(\omega)\operatorname{\mathbb{R}e}[\zeta(\omega)] \upd \omega
\end{equation}
The kinetic temperature that governs the width of the distribution of
the velocities of the heated particle follows from the equipartition
theorem \cref{eq:THBMk}, with $M$ replaced by the effective mass
$M_{\rm eff}$ which equals the mass of the particle $M$ augmented by
half the mass of the displaced fluid -- an effect that is solely the
consequence of the incompressible nature of the solvent and is made
apparent later.  To proceed further, we need the explicit form for
$\zeta(\omega)$ and we use the result for the friction coefficient on
a spherical colloidal particle in an incompressible solvent:\cite{Chow1973a}
\begin{equation}
  \label{eq:zeta_incompressible}
\zeta(\omega)=\zeta_0\left[ 1+a k_\omega (1-\complexi) -\frac{2
    \complexi}{9} a^2 k_\omega^2\right],
\end{equation}
where $\zeta_0=6 \pi \eta a$ is the Stokes friction coefficient. The
last term in the equation is responsible for the effective mass of the
particle. This becomes more evident once we write down the Green's
function as
\begin{equation}
  \label{eq:greens_function_1}
\begin{split}
G^{-1}(\omega)&=\zeta_0\left[1+ ak_\omega (1-\complexi) -\complexi\frac{2
    }{9} a^2 k^2_\omega -\complexi \frac{4}{9}
  \frac{\rho_P}{\rho_0} a^2 k_\omega^2\right]\\
&=\zeta_0\left[1+ ak_\omega (1-\complexi) -\complexi\frac{2
    }{9}\left(2\frac{\rho_P}{\rho_0}+1\right) a^2 k^2_\omega \right]
\end{split}
\end{equation}
where $\rho_P$ is the density of the colloid. It is apparent that the
above equation can simply be written down as 
\begin{equation}
  \label{eq:apparent_greens_function}
G^{-1}(\omega)=\zeta(\omega) - \complexi \omega  M_{\rm eff}
\end{equation}
with 
\begin{equation}
  \label{eq:effective_mass}
M_{\rm eff}=M+\frac{2}{3}\pi a^3 \rho_0=\frac{2}{3} \pi a^3 \rho_0
\left(2\frac{\rho_P}{\rho_0}+1 \right)
\end{equation}
but with $\zeta(\omega)=\zeta_0 [1+a k_\omega (1-\complexi)]$.
Plugging in the expression for $G(\omega)$ from
\cref{eq:greens_function_1} in \cref{eq:equal_time_correlation} and
using the variable $x=a k_\omega$ the kinetic temperature reads as
\begin{equation}
  \label{eq:kinetic_temperature}
\THBMk=\frac{1}{\pi}\int_0^\infty \frac{4\chi x(1+x) \mathcal{T}(x)
}{(1+x)^2 +x^2 (1+\chi x)^2} \upd x
\end{equation}
with $\chi=\frac{2}{9}(2 \rho_P/\rho_0+1)$.
\begin{figure}[!t]
  \centering
  \includegraphics[width=0.8\linewidth]{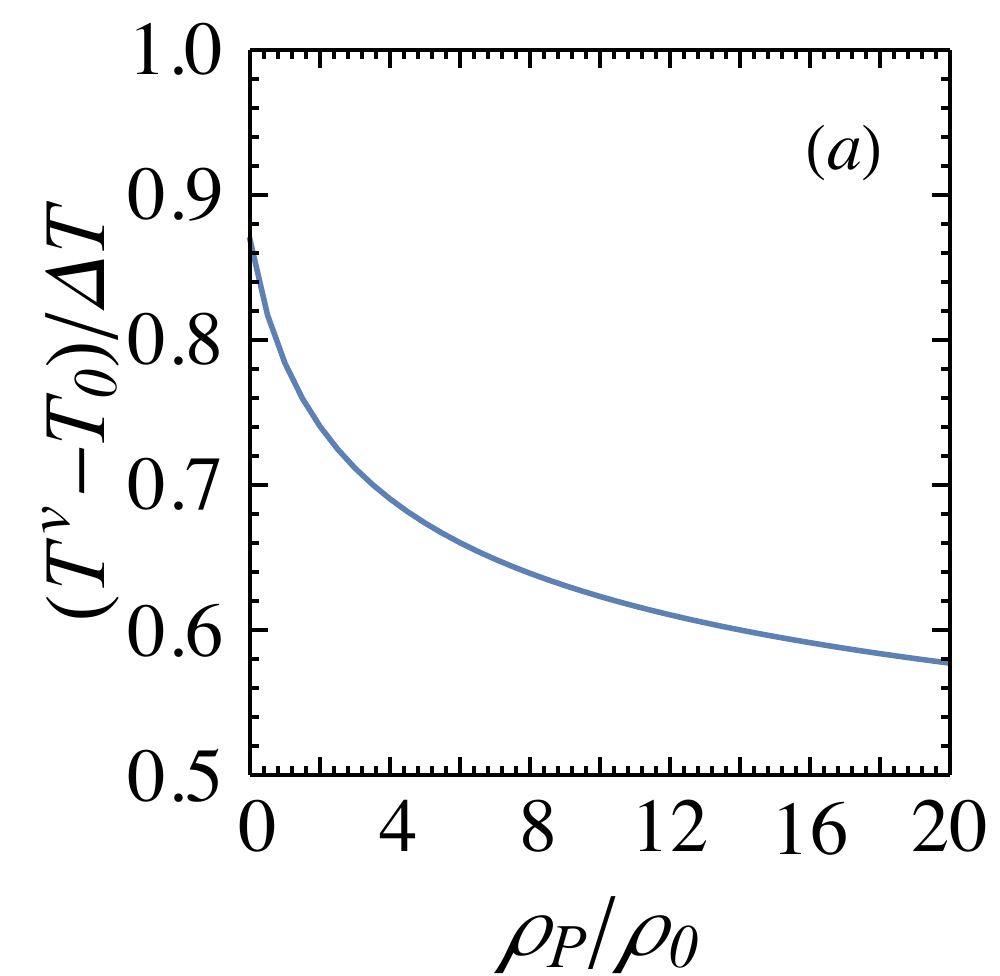}
  \caption{Plot of the kinetic temperature $\THBMk$ as a function of
    the density ratio.}
\end{figure}
Substituting for $\mathcal{T}(x)$ from \eqref{eq:T_omega}, we rewrite
the above equation as
\begin{equation}
  \label{eq:kinetic_temperature_1}
  \THBMk=T_0+\frac{5}{12} \Delta T\int_0^\infty 
\frac{4 \lambda x(1+x)\mathcal{F}(x)}{(1+x)^2+x^2 (1+\chi x)^2} \upd x
\end{equation}
The integral is numerically evaluated for different values of
$\chi$ and the quantity $\THBMk -T_0$ is plotted against the
density ratio $\rho_P/\rho_0$ for a better insight. We note that for
$\rho_P/\rho_0 \to 0$ such that the particle density $\rho_P$ remains
finite but the fluid density $\rho_0 \to \infty$ leads to an infinite
time for the vorticities to diffuse in the fluid and the frequency
dependent length scale $\lambda_\omega \to 0$. Consequently, the
particle equilibriates with the region of the fluid close to the
surface and the kinetic temperature
$\THBMk(\rho_P/\rho_0\to 0) \approx T_0+0.87 \Delta T$. The
opposite limit of $\rho_P/\rho_0 \to \infty$, with $\rho_0$ finite and
$\rho_P \to \infty$ is the Brownian limit where the particle is
infinitely massive. Due to this infinite inertia of the particle, the
kinetic degrees of freedom of the particle equilibriates with the
whole solvent and the kinetic temperature equals the configurational
temperature $\THBMk =\THBMc= T_0+\frac{5}{12}\Delta T$. 

\section{Frequency Dependent Diffusivity}
\label{sec:diffusivity}

One of the promising application of hot Brownian motion is its use in
Photothermal Correlation Spectroscopy (PhoCs) which has the advantage
of replacing the conventional fluorescence spectroscopic
techniques\cite{Radunz:2009ka}. Using hot Brownian motion to
investigate the local microscopic environment of complex fluids and
extract useful information about the viscoelastic response of such
systems would warrant an a-priori knowledge of the dependence of the
the diffusivity on the frequency arising from the inhomogeneous
temperature profile. By definition the frequency dependent diffusivity
is given by 
\begin{equation}
  \label{eq:diffusivity_def}
  D(\omega)=\frac{1}{3}\int_0^\infty \langle \vec{V}(t) \vec{V}(0) \rangle e^{\complexi
    \omega t} \upd t
\end{equation}
Using the form $\langle \vec{V}(t) \cdot \vec{V}(0) \rangle$ from
\cref{eq:vacf} the frequency dependent diffusion coefficient becomes
\begin{equation}
  \label{eq:diffusivity_freq_dep}
  D(\omega)=|G(\omega)|^2 \kb \mathcal{T}(\omega)\operatorname{\mathbb{R}e}[\zeta(\omega)]
\end{equation}
For $\omega=0$ the equation reduces to the value reported by Falasco
et. al \cite{Falasco:2014iq}: 
\begin{equation}
  \label{eq:effective_diffusivisty_omega0}
  D(0)=\frac{\kb \mathcal{T}(0)}{\zeta_0} =\frac{\kb T_0}{\zeta_0}
  \left(1+\frac{5}{12} \frac{\Delta T}{T_0} \right)
\end{equation}
indicating that the heated colloid moves with an enhanced diffusivity.
For frequencies close to $\omega=0$, the first order correction to
this value comes by expanding \eqref{eq:diffusivity_freq_dep} near
$\omega=0$ and leads to the expression:
\begin{equation}
  \label{eq:diffusion_coefficient_smal_frequency}
   D(\omega) \approx \frac{\kb T_0}{\zeta_0} \left[
  \left(1+\frac{5}{12} \frac{\Delta T}{T_0} \right)-a k_\omega
  \left(1+\frac{5}{6} \frac{\Delta T}{T_0} a k_\omega \right)\right]
\end{equation}
\section{Conclusion}
\label{sec:conclusion}
In conclusion, the framework of fluctuating hydrodynamics is used to
derive an expression for the frequency dependent effective temperature
$\mathcal{T}(\omega)$ of a hot Brownian particle. This effective
parameter is particularly useful in describing the dynamics of the
heated colloid using a generalized Langevin equation. We consider the
colloidal particle to be suspended in an incompressible solvent and
the subsequent heating of the colloid to be small. Consequently, the
resulting spatial variation of the viscosity profile can be neglected
in the linearized Stokes equation. Our work focuses on two specific
limits of this effective temperature - the low frequency limit
corresponding to $ \omega \to 0$ and the high frequency limit of
$\omega >>2 \eta/a^2 \rho_0$. 
In the low frequency limit, the first order correction to $\THBMc$ is
calculated and is show to scale as $\sqrt{\omega}$. 
 The analytical form $\mathcal{T}(\omega \to 0)$
and is further used in evaluating the effective frequency dependent
diffusion coefficient $D(\omega)$ and its approximation in the small
frequency limit. This is particularly useful in applications of hot
Brownian motion in experiments.
Further, a careful treatment of the corresponding high frequency limit
of $\mathcal{T}(\omega)$ is done by a suitable approximation of the
flow field for $\omega >> 2 \eta/ a^2 \rho_0 $. Finally,using the
generalized Lagevin equation and $\mathcal{T}(\omega)$, we estimate the
kinetic temperature $\THBMk$ that governs the distribution of the
velocities of the heated particle.  A numerical evaluation of the
equal time correlation $\langle \vec{V}^2 \rangle$ and using the
equipartition theorem yields $\THBMk$ as a function of ratio of the
densities of the particle and the fluid $\rho_P/\rho_0$ . A
consistency check is done by looking at the two limiting cases
$\rho_P/\rho_0 \to 0$ and $\rho_P/\rho_0 \to \infty$.

\appendix

\section{Calculation of the dissipation function}
\label{sec:dissipation_function_detail}

The terms $\mathcal{A}(r,\omega)$ and $\mathcal{B}(r,\omega)$ can be
further written down as
$\mathcal{A}(r,\omega)=\sum_{i=1}^3 a_i(r,\omega)$ and
$\mathcal{B}(r,\omega)=\sum_{i=1}^4 b_i(r,\omega)$, with
$a_i(r,\omega)$ given by
 \begin{equation}
\label{eq:ai}
\begin{split}
 a_1(r,\omega) &=\left( 1 -
   e^{-k_\omega(r-a)(1+\complexi)} \right)\\
a_2(r,\omega) &=\frac{2}{3} \complexi a^2 k_\omega^2 \left( 1-
  \frac{r^2}{a^2} e^{-k_\omega(r-a)(1+\complexi)} \right)\\
a_3(r,\omega) &= (1+\complexi)a k_\omega \left( 1- \frac{r}{a}e^{-k_\omega(r-a)(1+\complexi)}\right)
\end{split}
 \end{equation}
and $b_i(r,\omega)$ 
 \begin{equation}
\label{eq:bi}
\begin{split}
 b_1(r,\omega) &=\left( 1 -
   e^{-k_\omega(r-a)(1+\complexi)} \right)\\
b_2(r,\omega) &=\frac{2}{3} \complexi a^2 k_\omega^2 \left( 1-
  \frac{3}{2}\frac{r^2}{a^2} e^{-k_\omega(r-a)(1+\complexi)} \right)\\
b_3(r,\omega) &= (1+\complexi)a k_\omega \left( 1-
  \frac{r}{a}e^{-k_\omega(r-a)(1+\complexi)}\right) \\
b_4(r,\omega) &=-\complexi (1+\complexi) a^3 k_\omega^3 \frac{r^3}{a^3}e^{-k_\omega(r-a)(1+\complexi)}.
\end{split}
 \end{equation}
Using the above equations, the expressions for
$\mathcal{A}\mathcal{A}^*$ and $\mathcal{B}\mathcal{B}^*$ can be
recast in the form 
\begin{equation}
\label{eq:aa*}
\begin{split}
  \mathcal{A}\mathcal{A}^*=g_1(k_\omega)&+g_2(r,k_\omega)e^{-k_\omega
    (r-a)} \sin k_\omega (r-a) +\\
& g_3(r,k_\omega)e^{-k_\omega
    (r-a)} \cos k_\omega (r-a) +\\
&\quad \quad \quad \quad g_4(r,k_\omega)e^{-2k_\omega
    (r-a)}
\end{split}
\end{equation}
and
\begin{equation}
\label{eq:bb*}
\begin{split}
  \mathcal{B}\mathcal{B}^*=f_1(k_\omega)&+f_2(r,k_\omega)e^{-k_\omega
    (r-a)} \sin k_\omega (r-a) +\\
& f_3(r,k_\omega)e^{-k_\omega
    (r-a)} \cos k_\omega (r-a) +\\
&\quad \quad \quad \quad f_4(r,k_\omega)e^{-2k_\omega
    (r-a)},
\end{split}
\end{equation}
where the coefficients $g_i$ reads:
\begin{equation}
\label{eq:g_i}
\begin{split}
  g_1(k_\omega) &=1+ 2 a k_\omega + 2 a^2 k_\omega^2 +\frac{4}{3} a^3
  k_\omega^3 +\frac{4}{9} a^4 k_\omega^4 \\
g_2(r,k_\omega) &=\left(1+ 2 k_\omega r +2 k_\omega^2 r^2 +\frac{4}{3}
k_\omega^3 r^3 + \frac{4}{9} k_\omega^4 r^4 \right)\\
g_3(r,k_\omega)&=-2\left(1+a k_\omega +r k_\omega +2 a k_\omega^2 r +
\frac{2}{3} a^2 k_\omega^3 r \right.\\
& \left. \quad \quad \quad +\frac{2}{3} a k_\omega^3 r^2
+\frac{4}{9} a^2 k_\omega^4 r^2 \right) \\
g_4(r,k_\omega) &= 2 a k_\omega \left(1 + \frac{2}{3} a k_\omega -
  \frac{r}{a} + \frac{2}{3} a k^2_\omega r   \right. \\
& \left. \quad \quad \quad \quad \quad \quad- 2 a k_\omega
  \frac{r^2}{a^2} -\frac{2}{3} k^2_\omega  r^2 \right)
\end{split}
\end{equation}
and the coefficients $f_i$ are given by 
\begin{equation}
\label{eq:f_i}
\begin{split}
  f_1(k_\omega) &=1+ 2 a k_\omega + 2 a^2 k_\omega^2 +\frac{4}{3} a^3
  k_\omega^3 +\frac{4}{9} a^4 k_\omega^4 \\
  f_2(r,k_\omega) &=\left(1+ 2 k_\omega r +2 k_\omega^2 r^2
    +\frac{4}{3}
    k_\omega^3 r^3 \right. \\
  &\left. \quad \quad \quad \quad \quad \quad \quad
    + k_\omega^4 r^4 + \frac{2}{3} k_\omega^5 r^5 + \frac{2}{9} k_\omega^6 r^6 \right)\\
  f_3(r,k_\omega)&=-2\left(1+a k_\omega +r k_\omega +2 a k_\omega^2 r
    +
    \frac{2}{3} a^2 k_\omega^3 r \right.\\
  & \left. + a k_\omega^3 r^2 +\frac{2}{3} a^2 k_\omega^4 r^2 -
    \frac{1}{3} k_\omega^3 r^3
    +\frac{2}{9} a^2 k_\omega^5 r^3\right) \\
  f_4(r,k_\omega) &= 2 a k_\omega \left(1 + \frac{2}{3} a k_\omega -
    \frac{r}{a} + \frac{2}{3} a k^2_\omega r - a k_\omega \frac{r^2}{a^2} \right. \\
  & \left. \quad - k^2_\omega r^2 - \frac{1}{3} a^2
    k^2_\omega\frac{r^3}{a^3} -\frac{2}{3} k^3_\omega r^3 -
    \frac{2}{9} a k_\omega^4 r^3 \right)
\end{split}
\end{equation}

\section{Calculation of the effective temperature in the high
  frequency limit}
\label{sec:Teff_high_frequency_limit}
Using the expression for $\mathcal{A}_1$ and $\mathcal{B}_1$ from \cref{eq:A_1_approx,eq:B_1_approx}, we
rewrite the dissipation function as 
\begin{equation}
  \label{eq:dissipation_function_high_frequency_approx}
\begin{split}
\phi(\vec{r},\omega)=2 \eta |U_\omega|^2 \left( \frac{81 a^2 \eta^2}{\rho^2_0
    \omega^2}\right)\frac{1}{r^8} \bigg[
3 f_1(k_\omega) \cos^2 \theta +\\
\tilde{g}(r,k_\omega)\sin^2 \theta  \bigg]
\end{split}
\end{equation}
where $f_1(k_\omega)$ is defined in \cref{eq:f_i} and
$\tilde{g}(r,k_\omega)$ is given by
\begin{equation}
\begin{split}
  \tilde{g}(r,k_\omega) &=1+ 2 a k_\omega + 2 a^2 k_\omega^2 +\frac{4}{3} a^3
  k_\omega^3 +\frac{4}{9} a^4 k_\omega^4 \\
&+2 k^3_\omega r^3
  e^{-k_\omega (r-a)}-\frac{4}{3} a^2 r^3 k^5_\omega
  e^{-k_\omega(r-a)} \\
&+2 k^6_\omega r^6 e^{-2k_\omega (r-a)}
\end{split}
\end{equation}
The spatial integral of the dissipation function yields 
\begin{equation}
\label{eq:dissipation_function_integral_final}
\int \upd \vec{r} \phi(\vec{r},\omega)=2 \eta |U_\omega|^2 \left( \frac{81 a^2 \eta^2}{5\rho^2_0
    \omega^2}\right)\left(\frac{1}{a^5}\right)\mathcal{G}(\omega)
\end{equation}
with
\begin{equation}
\label{eq:G_omega}
\mathcal{G}(k_\omega)=\bigg[2 f_1(k_\omega) +\frac{4}{27} \tilde{g}_1(k_\omega)\bigg]
\end{equation}
where $\tilde{g}_1(k_\omega)$ is given by
\begin{equation}
\begin{split}
  \tilde{g}_1(k_\omega)&=\frac{1}{45 a^5}\bigg(9+ 18 a k_\omega +18 a^2
    k^2_\omega+57 a^3 k^3_\omega\\
&-41 a^4 k^4_\omega +15 a^5 k^5_\omega
    +30 a^6 k^6_\omega \\
&+15 a^5 e^{a k_\omega} k_\omega^5 \left(-3+2
      a^2 k_\omega ^2\right) \text{Ei}(-ak_\omega)\bigg)
\end{split}
\end{equation}
The numerator in the expression for the effective temperature
evaluates to 
\begin{equation}
  \label{eq:numerator_high_frequency}
\begin{split}
  \int \upd \vec{r} T(r) \phi(\vec{r},\omega)=2T_0 \eta |U_\omega|^2
  \left( \frac{81 a^2 \eta^2}{5\rho^2_0
      \omega^2}\right)\left(\frac{1}{a^5}\right)\mathcal{G}(\omega)\\
  +2 \eta |U_\omega|^2 \left( \frac{81 a^2 \eta^2}{6\rho^2_0
      \omega^2}\right)\left(\frac{1}{a^6}\right)\mathcal{H}(\omega)
\end{split}
\end{equation}
with $\mathcal{H}(\omega)$ given by
\begin{equation}
\label{eq:H_omega}
\mathcal{H}(k_\omega)=\bigg[2 f_1(k_\omega) +\frac{4}{27} \tilde{g}_2(k_\omega)\bigg]
\end{equation}
where $\tilde{g}_2(k_\omega)$ has the form
\begin{equation}
  \label{eq:g2_omega}
\begin{split}
\mathcal{H}(\omega)&=\bigg[9+18 a k_\omega+18 a^2 k_\omega^2+48 a^3
k_\omega^3-14 a^4 k_\omega^4\\
&-6 a^5 k_\omega^5+12 a^6 k_\omega^6-12 a^7 k_\omega^7\\
&-6 a^6 k_\omega^6 e^{a k_\omega} \left(-3+2 a^2 k_\omega^2\right)
\text{Ei}(-a k_\omega)\\
&+108 a^6 k_\omega^6 e^{2 a k_\omega} \Gamma(0,2 a k_\omega)\bigg],
\end{split}
\end{equation}
where $\Gamma \left(0, x \right) = \int\limits_0^\infty {s^{x - 1} e^{ - s} ds}$.

\bibliography{kinetic_temperature}

\end{document}